# THEORETICAL FOUNDATIONS AND COVARIANT BALANCES FOR CHEMICAL ENGINEERING APPLICATIONS WITH ELECTROMAGNETIC FIELD

***Jean-François CORNET***

***Laboratoire de Génie Chimique et Biochimique***

***Université Blaise Pascal. Bât. CUST. 24, Avenue des Landais, BP 206. F-63174 AUBIERE Cedex***

***Phone: 33 *4 73 40 50 56. Fax: 33 *4 73 40 78 29***

***J-Francois.CORNET@univ-bpclermont.fr***

**Running title**: Covariant Balances with Electromagnetic Field

## Abstract

A covariant formalism is used in order to examine the status of Maxwell equations and to unify the concept of balances, for all chemical engineering applications in relation with electrodynamics. The resulting formal structure serves as a discussion in studying the determinism of electromagnetic systems, and examining the theoretical foundations for a general classification of chemical engineering applications when non-conservative forces are concerned. A strategy for modeling such applications is then sketched and the balances for the main conserved and non-conserved extensive quantities are then summarized in their covariant and classical forms.

## Introduction

Electrodynamics (in the broadest sense) is a field of increasing importance in relation with chemical engineering applications, both in conventional industry or in high technology sectors. Modeling and scheduling such applications requires to properly establish a general formalism in which balances are formulated in order to verify the invariance principles [1-3]. Because they play a crucial role in electromagnetic problems, this objective also involves to discuss the status of Maxwell equations. These equations, containing the velocity of light as an intrinsic parameter are obviously available in special and general relativity. Consequently, a formal structure of balances can only be reached in a four dimensional space-time continuum, requiring a general covariant formulation. This ensures the invariance of any physical law under Lorentz transformation which then appears as an orthogonal transformation in the Minkowski space, satisfying *per se* the equivalence principle [2].

The aim of this paper is to examine how the well-established general covariant four-dimensional approach is able to integrate the classical local balances [4-6] together with the well-known Maxwell equations in an unified point of view. In relation, the status of Maxwell equations as covariant balances is investigated. The resulting set of equations, necessary to physically formulate a chemical engineering problem in relation to electrodynamics, provides a theoretical basis in order to investigate the mathematical foundations of such applications. This leads to a consistent analysis of the coupling between the material and the electromagnetic phases of the process, enabling to propose a general classification of chemical engineering problems, both for electrodynamics and wave transport phenomena.

**Classical Balances for Inertial and Galilean Frame of Reference.**

Different ways exist in order to formulate local balances for classical applications, with only conservative forces. Authors have used the invariance principle in a galilean frame of motion [3], conservation principles [5], differential balances [4] or more axiomatic approaches [6]. An alternative well-known approach, which presents the advantage to be quite general, is briefly presented in the following.

One considers a given extensive quantity, which may be a scalar or a vector $\Psi_{(i)}$ (the subscript *i* between brackets authorizes that $\Psi$ be a vector) and flowing through a volume element *dV*, defined by a moving control surface *F*. $\psi_{(i)}$ corresponds to the quantity $\Psi_{(i)}$ by unit volume and $J_{(ij)}$ is the flux density of $\Psi_{(i)}$, considering an outward normal to the system (the second subscript *j* authorizes that *J* be a tensor, the rank of *J* being always one order higher than $\Psi_{(i)}$). Actually, *J* is the sum of three terms :

$$J_{(ij)} = J_{(ij)l} - J_{(ij)f} + J_{(ij)c} = J_{(ij)r} + J_{(ij)c} \qquad (1)$$

where $J_{(ij)l}$ is a convective flux density ; $J_{(ij)f}$ is a flux density related to the moving frontier ( $J_{(ij)f} = \psi_{(i)} v_{(j)f}$ ) ; $J_{(ij)c}$ is a conductive flux density. Finally, $J_{(ij)r} = J_{(ij)l} - J_{(ij)f}$ is a flux density by relative displacement.

The term $J_{(ij)r}$ corresponds to the transport of the flux density of $\Psi_{(i)}$ by the material phase, equal to $\dot{\Psi}_{(i)} dw_m$ where $\dot{\Psi}_{(i)}$ is the quantity $\Psi_{(i)}$ by unit mass, and $dw_m = -\rho \boldsymbol{v}_r \cdot d\boldsymbol{S}$ . It can also include forced convection terms by inter-phase mass transfer. The last term $J_{(ij)c}$ corresponds to the conductive flux density at the right of fluid sections, locally of major importance. Finally, it is necessary to take into account a source term of $\Psi_{(i)}$ for non-conserved quantities, which is $\pm\sigma_{(i)}$ .

***Spatial and Local Balances***

The general spatial form for the balance of the quantity $\Psi$ is then :

$$\frac{d\Psi_{(i)}}{dt} = -\iint_S J_{(ij)} \dot{\times} d\boldsymbol{S} + \iiint_V \sigma_{(i)} dV \qquad (2)$$

where the operator " $\dot{\times}$ " involves the vectorial or tensorial nature of *J*. This formulation corresponds to a generalization of the conservation laws of mechanics, but does not contain any information about the terms $J_{(ij)}$ and $\sigma_{(i)}$. From eq. (1) one has for the spatial balance :

$$\frac{d\Psi_{(i)}}{dt} = -\iint_S J_{(ij)l} \dot{\times} d\boldsymbol{S} + \iint_S J_{(ij)f} \dot{\times} d\boldsymbol{S} - \iint_S J_{(ij)c} \dot{\times} d\boldsymbol{S} + \iiint_V \sigma_{(i)} dV \qquad (3)$$

The local balance formulation requires to develop the time derivative of $\Psi_{(i)}$ :

$$\frac{d\Psi_{(i)}}{dt} = \frac{d}{dt} \iiint_V \psi_{(i)} dV \qquad (4)$$

in which we have $\psi_{(i)} = \rho \dot{\Psi}_{(i)}$ only for a material phase. One can then expand the eq. (3) in the form :

$$\frac{d\Psi_{(i)}}{dt} = \frac{d}{dt} \iiint_V \psi_{(i)} dV = \iiint_V \frac{\partial \psi_{(i)}}{\partial t} dV + \iiint_V \psi_{(i)} \frac{d}{dt}(dV) \qquad (5)$$

Because the displacement of the frontier $F$ is the sole way to cause a volume variation $dV$, one has : $d(dV) = \boldsymbol{dS} \cdot \boldsymbol{v}_f dt$ and the corresponding flux of $\Psi_{(i)}$ then is :

$$\iiint_V \psi_{(i)} \frac{d}{dt}(dV) = \iint_S \psi_{(i)} v_{(j)f} \dot{\times} \boldsymbol{dS} = \iint_S J_{(ij)f} \dot{\times} \boldsymbol{dS} \quad (6)$$

which enables to write eq. (5) in the general form :

$$\frac{d}{dt}\iiint_V \psi_{(i)} dV = \iiint_V \frac{\partial \psi_{(i)}}{\partial t} dV + \iint_S J_{(ij)f} \dot{\times} \boldsymbol{dS} \quad (7)$$

This relation is known as the Leibnitz rule or generalized transport theorem [1,5,6] and it is a purely kinematics equation, available even for a non-material phase.

The identification with the terms of the spatial balance (eq. 3) gives :

$$\iiint_V \frac{\partial \psi_{(i)}}{\partial t} dV = -\iint_S (J_{(ij)l} + J_{(ij)c}) \dot{\times} \boldsymbol{dS} + \iiint_V \sigma_{(i)} dV \quad (8)$$

and using the divergence theorem :

$$\iint_S J_{(ij)} \dot{\times} \boldsymbol{dS} = \iiint_V \frac{\partial}{\partial x_i} J_{(ij)} dV$$

one obtains :

$$\iiint_V \left[ \frac{\partial \psi_{(i)}}{\partial t} + \frac{\partial}{\partial x_i}(J_{(ij)c} + J_{(ij)l}) - \sigma_{(i)} \right] dV = 0 \quad (9)$$

Because the frontier $F$ is arbitrarily defined, the integral vanishes, giving the local balance of $\Psi_{(i)}$ :

$$\frac{\partial \psi_{(i)}}{\partial t} = -\frac{\partial}{\partial x_i}(J_{(ij)c} + J_{(ij)l}) + \sigma_{(i)} \quad (10)$$

This equation corresponds to the general eulerian form for an extensive quantity balance in an inertial and galilean frame of reference [7, 8], and leads to the famous equations of change [4, 5] for chemical and mechanical engineering applications. A more rigorous demonstration of this equation is available in the relevant literature [1, 6, 7, 9]. The non-inertial form of eq. (10) is further discussed in the textbooks of Dennery [8] and Wilmansky [6].

**Covariant Formalism for Inertial and Relativistic Frame of Reference**

As previously explained, the classical approach excludes modeling of chemical engineering applications in which external forces are non-conservative (the long range volumetric forces acting on the system, and their corresponding work, appearing in the source term of the material balances are assumed conservative), i.e. electrodynamics applications. In these cases, it is necessary to use the well-known Maxwell equations, the formal structure of which is not clearly debated, except by some authors [10]. From the physical point of view, Maxwell equations which intrinsically contain the speed of light, may appears as principles, valid in special and general relativity. Consequently, their integration in a general balances formalism must be sought for only in a Minkowski four-dimensions space-time frame of reference, even for chemical engineering applications in which the non-relativistic assumption is always valid.

The reader is here assumed familiar with tensorial notations in four-dimensions space-time frame of reference [10, 11].

The 4-vectors and covariant 4-pseudotensors are given in the form $A_\lambda = (-a, \boldsymbol{A})$ and $A_{\lambda\mu}$ (scalar capacity); the 4-vectors and contravariant 4-pseudotensors are $A^\lambda = (-\frac{a}{c^2}, -\boldsymbol{A})$ and $A^{\lambda\mu}$ (scalar density). The covariant and contravariant quantity are related by the metric tensor $g^{\lambda\mu}$:

$$g^{\lambda\mu} = g_{\lambda\mu} = \begin{bmatrix} c^2 & 0 & 0 & 0 \\ 0 & -1 & 0 & 0 \\ 0 & 0 & -1 & 0 \\ 0 & 0 & 0 & -1 \end{bmatrix} \text{ with } g_{\lambda\mu}A^\mu = A_\lambda \text{ and } g^{\lambda\mu}A_\mu = A^\lambda$$

From these definitions, one can examine the re-formulation of balances in the form of eq. (10), in order to generalize the approach to non-conservative forces, i.e. taking into account the electromagnetic field. The problem is then quite different for a scalar or a vector extensive quantity.

***Scalar Balances***

In this case, the flux density is unambiguously and easily defined by the 4-vector $J^{\mu}_{c+l} = (\psi, \boldsymbol{J}_{c+l})$, leading to write the four-dimensional form of eq. (10) as :

$$\frac{\partial J^{\mu}_{c+l}}{\partial x^{\mu}} = \sigma \qquad (11)$$

which is the local balance of the quantity $\Psi_{(i)}$ in the Minkowski space, and restores, in a galilean space a scalar equation in divergence and $\partial/\partial t$.

***Vector Balances***

In this case, the flux density is a 4-tensor, and eq. (10) takes the general form :

$$\frac{\partial J^{\lambda\mu}_{c+l}}{\partial x^{\mu}} = \sigma^{\lambda} \qquad (12)$$

where $\sigma^{\lambda}$ is a 4-vector source. Dramatically, the 4-tensor $J^{\lambda\mu}_{c+l}$ can have different forms, depending if it is related to a material phase (momentum balance for example) or to an electromagnetic extensive quantity.

In the case of a material phase, the 4-tensor $J^{\lambda\mu}_{c+l}$ is a true symmetric tensor. For the non-relativistic approximation, it takes the following form [12] :

$$J^{\lambda\mu}_{c+l} = \begin{bmatrix} J^{00} & \psi_x & \psi_y & \psi_z \\ \psi_x & J_{(xx)c+l} & J_{(xy)c+l} & J_{(xz)c+l} \\ \psi_y & J_{(yx)c+l} & J_{(yy)c+l} & J_{(yz)c+l} \\ \psi_z & J_{(zx)c+l} & J_{(zy)c+l} & J_{(zz)c+l} \end{bmatrix} \qquad (13)$$

This 4-tensor is then structured in three different terms : a scalar $J^{00} = J_{00}$, a 3-vector $J^{0\mu} = \boldsymbol{\psi}$ corresponding to a volumetric density, and a 3-tensor $J_{(ij)c+l} = J_{(ji)c+l}$ corresponding to the convective and conductive flux density. Clearly, applying eq. (13) to non-relativistic momentum balance gives : $J^{00} = \rho\dot{U} + \frac{1}{2}\rho v^2$, energy density for the fluid, $\psi_i = \rho \boldsymbol{v}$, and $J_{(ij)c+l} = \Pi_{ij} + \rho v_i v_j$ with $\Pi_{ij} = \tau_{ij} + P\delta_{ij}$. From this definition, the general four-dimensional equation (12) restores, from the spatial coordinates, the well-known momentum balance for a material phase [4,5,12] ; the temporal coordinate giving a colinear equation. Of course, it must be emphasized that this approach does not present any interest for classical chemical engineering applications on a material phase, because the momentum balance can be easily obtained from the classical way, using eq. (10) ; it is just here a formal presentation to keep some generality to the proposed approach.

For extensive electromagnetic quantity balances, the flux density terms are assimilated to conductive term, because obviously independent of the material phase. Moreover, the 4-tensor flux density term is in reality an anti-symmetric 4-pseudotensor (a scalar density), or a six-vector defined as :

$$J_c^{\lambda\mu} = \begin{bmatrix} 0 & \psi_x & \psi_y & \psi_z \\ -\psi_x & 0 & J_{(z)c} & -J_{(y)c} \\ -\psi_y & -J_{(z)c} & 0 & J_{(x)c} \\ -\psi_z & J_{(y)c} & -J_{(x)c} & 0 \end{bmatrix} \quad (14)$$

$\boldsymbol{\psi}$ is a polar vector and $\boldsymbol{J}_c$ is an axial vector. A partial mathematical demonstration of the structure of this 4-tensor is available in the related literature [10, 11, 13]. From this definition, the four-dimensional equation (12) restores in a galilean frame of reference, a vector equation in curl and $\partial/\partial t$ terms with a scalar equation in divergence, leading to the two non-homogeneous Maxwell equations, if applied to the electromagnetic fields.

The major difference of structure between equations (13) and (14) comes from the fact that for a material phase, the balance is obtained considering a motion in a constant field, whereas for electromagnetic field balance one considers a given motion in varying the field potentials which then take place for “coordinates”.

### *Application to Balances in an Electromagnetic Field*

As a first example, using equation (11) in the field of electrodynamics, it is easy to show how the charge continuity equation for the material phase is obtained. Defining the current density 4-vector by $J_{c+l}^{\mu} = (\rho_e, \boldsymbol{J}_{c+l}) = (J^0, \boldsymbol{J}_{c+l})$ and taking $\sigma = 0$ for the charge conservation, eq. (11) immediately gives :

$$\frac{\partial J_{c+l}^{\mu}}{\partial x^{\mu}} = \frac{\partial \rho_e}{\partial t} + \frac{\partial J_x}{\partial x} + \frac{\partial J_y}{\partial y} + \frac{\partial J_z}{\partial z} = \frac{\partial \rho_e}{\partial t} + div\boldsymbol{J}_{c+l} = 0 \quad (15)$$

which is indeed a charge balance $\Psi = Q$ for the material phase, also given by eq. (10). At the contrary, if the second example deals with extensive quantities of the electromagnetic fields, only equation (12) is concerned. The 4-tensor of electric and magnetic excitations then takes the form of eq. (14) :

$$J_c^{\lambda\mu} = G^{\lambda\mu} = \begin{bmatrix} 0 & G^{01} & G^{02} & G^{03} \\ G^{10} & 0 & G^{12} & G^{13} \\ G^{20} & G^{21} & 0 & G^{23} \\ G^{30} & G^{31} & G^{32} & 0 \end{bmatrix} = \begin{bmatrix} 0 & D_x & D_y & D_z \\ -D_x & 0 & H_z & -H_y \\ -D_y & -H_z & 0 & H_x \\ -D_z & H_y & -H_x & 0 \end{bmatrix} \quad (16)$$

and the 4-vector source is the current density $J^{\lambda} = (\rho_e, \boldsymbol{J})$, leading to the balance, from eq. (12), [10, 13] :

$$\frac{\partial G^{\lambda\mu}}{\partial x^{\mu}} = J^{\lambda} \quad (17)$$

which is the Minkowski's form of the two non-homogeneous Maxwell equations. Taking the three spatial coordinates of eq. (17), $\lambda = 1, 2, 3$, one has indeed :

$$\begin{cases} \partial_0 G^{10} + \partial_1 G^{11} + \partial_2 G^{12} + \partial_3 G^{13} = J^1 \\ \partial_0 G^{20} + \partial_1 G^{21} + \partial_2 G^{22} + \partial_3 G^{23} = J^2 \\ \partial_0 G^{30} + \partial_1 G^{31} + \partial_2 G^{32} + \partial_3 G^{33} = J^3 \end{cases} \text{ that is : } \operatorname{curl} \boldsymbol{H} - \frac{\partial \boldsymbol{D}}{\partial t} = \boldsymbol{J} \quad (18)$$

and taking the temporal coordinate, gives :

$$\partial_0 G^{00} + \partial_1 G^{01} + \partial_2 G^{02} + \partial_3 G^{03} = J^0 \text{, or } \operatorname{div} \boldsymbol{D} = \rho_e \quad (19)$$

Clearly, the balance equation (17) leads to the first pair of Maxwell equations (eq. 18 and 19) corresponding respectively to extensive quantities defined by the local magnetic moment $\boldsymbol{\Psi} = \boldsymbol{m} = \mathcal{M}V (\text{or } \boldsymbol{j} = \mathcal{J}V)$, and the local electric moment $\boldsymbol{\Psi} = \boldsymbol{p} = \mathcal{P}V$ and appearing in the thermodynamic treatment of electromagnetic systems [14].

The main balances for chemical engineering applications, in classical three-dimensional space, and in the Minkowski form in a four-dimensional space, are summarized in Table 1. A proof in obtaining the balance equation (17) using the lagrangian formalism and the least action principle is also given in Appendix 1.

**Covariant Formalism for Electromagnetic Field**

The first part of this paper dealt with classical applications, with constant fields deriving from a scalar potential $\operatorname{grad} \dot{\Phi}$ with $\dfrac{\partial \dot{\Phi}}{\partial t} = 0$ (as gravity $\boldsymbol{g}$ and electric field $\boldsymbol{E}$ for example). The field equation is often omitted in classical formulation of problems (definition of gravity) whereas it is a necessary relation to obtain the internal energy balance from the total energy conservation equation. In case of electromagnetic applications, it is necessary to extend the definition to variable electric and magnetic fields, which is classically feasible defining the vector potential $\boldsymbol{A}$ :

$$\boldsymbol{E} = -\operatorname{grad} V - \frac{\partial \boldsymbol{A}}{\partial t} \quad (20) \qquad \text{and} \qquad \boldsymbol{B} = \operatorname{curl} \boldsymbol{A} \quad (21)$$

Taking the curl of eq. (20) and the divergence of eq. (21), then applying the identities curl grad = 0 and div curl = 0, leads immediately to the two homogeneous Maxwell equations :

$$\operatorname{div} \boldsymbol{B} = 0 \quad (22)$$

$$\operatorname{curl} \boldsymbol{E} + \frac{\partial \boldsymbol{B}}{\partial t} = 0 \quad (23)$$

which then appear as variable fields definition equations and not as balance equations.

In order to keep an intrinsic formal structure to this presentation, it is possible to obtain the Minkowski formulation of these two equations in a four-dimensional frame, from the definition of the covariant 4-tensor for fields $F_{\lambda\mu}$:

$$F_{\lambda\mu} = \frac{\partial A^{\nu}}{\partial x^{\lambda}} - \frac{\partial A^{\lambda}}{\partial x^{\nu}} = \operatorname{curl} A^{\mu} = \begin{bmatrix} 0 & -E_x & -E_y & -E_z \\ E_x & 0 & B_z & -B_y \\ E_y & -B_z & 0 & B_x \\ E_z & B_y & -B_x & 0 \end{bmatrix} \quad (24)$$

from which the field equations (22-23) may be condensed in [10] :

$$\frac{\partial F_{\nu\rho}}{\partial x^{\mu}} + \frac{\partial F_{\rho\mu}}{\partial x^{\nu}} + \frac{\partial F_{\mu\nu}}{\partial x^{\rho}} = 0 \ \text{or,} \ \partial_{[\mu} F_{\nu\rho]} = 0 \quad (25)$$

Once more, this equation is not a balance, but a relation between fields and the generalized vector potential. Nevertheless, it is necessary to formulate all the electrodynamics relevant applications. So, it appears as fields definition in Table 1.

**Table 1 :** **Minkowski covariant form and classical form for the main balances of interest when chemical engineering applications with electromagnetic field are concerned.**

| **Balance** | ***Covariant Formulation*** <br> ***(Minkowski formalism)*** | ***Classical Formulation*** <br> ***(Galilean formalism)*** |
|---|---|---|
| ***Mass*** | $\frac{\partial J^{\mu}}{\partial x^{\mu}} = 0 \qquad [\, J^{\mu} = (\rho_{(e)}, \boldsymbol{J})\,]$ <br> (conserved) | $\frac{\partial \rho_{(e)}}{\partial t} = -\operatorname{div} \boldsymbol{J}$ |
| ***Momentum*** | $\frac{\partial T^{\lambda\mu}}{\partial x^{\mu}} = 0$ <br> $T^{\lambda\mu}$ is the energy-momentum tensor [11] <br> (conserved) | $\frac{\partial}{\partial t}\left[\rho \boldsymbol{v} + (\boldsymbol{D} \times \boldsymbol{B})\right] = -\frac{\partial}{\partial x_i}\left[\rho v_i v_j + \Pi_{ij} - T_{ij}\right]$ <br> where the tensor $\Pi_{ij} = \tau_{ij} + P\delta_{ij}$ |
| ***Total Energy*** | $\frac{\partial E^{\mu}}{\partial x^{\mu}} = 0 \qquad [\, E^{\mu} = (E_{TOT}, \boldsymbol{J}_{E_{TOT}})\,]$ <br> (conserved) | $\frac{\partial E_{TOT}}{\partial t} = -\operatorname{div}\begin{bmatrix} \boldsymbol{S} + \frac{1}{2}\rho v^2 \boldsymbol{v} + \tau_{ij} \times \boldsymbol{v} + \rho \dot{U}\, \boldsymbol{v} + \sum_k \dot{\Phi}_k \boldsymbol{J}_k \\ + P\boldsymbol{v} - (\boldsymbol{E} \cdot \boldsymbol{\mathcal{P}})\boldsymbol{v} - (\boldsymbol{H} \cdot \boldsymbol{\mathcal{J}})\boldsymbol{v} + \boldsymbol{J}_Q \end{bmatrix}$ <br> where $E_{TOT} = w + \frac{1}{2}\rho v^2 + \rho \dot{U} + \rho_{(e)} \dot{\Phi}$ |
| ***Electromagnetic Moment*** <br> *(non-homogeneous Maxwell equations)* | $\frac{\partial G^{\lambda\mu}}{\partial x^{\mu}} = J^{\lambda}$ <br> (not conserved) | $\operatorname{div} \boldsymbol{D} = \rho_e$ <br> $\operatorname{curl} \boldsymbol{H} - \frac{\partial \boldsymbol{D}}{\partial t} = \boldsymbol{J}$ |
| **Electromagnetic Field** <br> *(homogeneous Maxwell equations)* | $\frac{\partial F_{\nu\rho}}{\partial x^{\mu}} + \frac{\partial F_{\rho\mu}}{\partial x^{\nu}} + \frac{\partial F_{\mu\nu}}{\partial x^{\rho}} = 0$, or $\partial_{[\mu} F_{\nu\rho]} = 0$ | $\operatorname{div} \boldsymbol{B} = 0$ or $\boldsymbol{B} = \operatorname{curl} \boldsymbol{A}$ <br> $\operatorname{curl} \boldsymbol{E} + \frac{\partial \boldsymbol{B}}{\partial t} = 0$ or $\boldsymbol{E} = -\operatorname{grad} V - \frac{\partial \boldsymbol{A}}{\partial t}$ |

**Determinism of Electromagnetic Systems**

Writing balance equations to entirely determine and formulate a chemical engineering problem (or in the broadest sense an engineering problem in continua) is an important question already debated [3-5,14-16]. The extension of this problem to applications in electrodynamics of continua is here of prime importance. Classical textbooks present different approaches in relation with this question [2, 11]. In its two major monographs Truesdell [1, 17], mentions the six main conservative quantities for this kind of problems : the mass, the momentum, the moment of momentum, the total energy, the charge and the magnetic flux, arguing that this enables to generalize the mechanics of continua to the electrodynamics of continua. As previously discussed in this paper, this appears as a reduced view of the problem, because modeling most of electrodynamics applications requires to use the four (or at least two) Maxwell equations [2, 18]. Moreover, it is easy to show that the conservation of charge is already contained in the Maxwell equations, i.e. it is a redundant equation (Appendix 2). Consequently, one can conclude that the unified treatment of electrodynamics applications requires to use, in addition to the four classical Minkowski balances (mass, momentum, moment of momentum and energy), only the Minkowski electromagnetic moment balance (eq. 17). Nevertheless, it must be emphasized that any electromagnetic problem formulation requires also to use equations defining the non-stationary fields, and that a complete description often requires to explain internal energy or entropy balances, which is impossible without a thermodynamic treatment (second principle, Gibbs equation - see [3, 14, 15] -). Finally, a considerable difficulty to transform a balance equation into a more tractable field of intensive quantity equation may result in defining constitutive relations characterizing the material phase, specially for non-linear media [10, 15, 19].

In addition to their general form given in Table 1, the main balances discussed above are summarized in their practical form in Table 2, both on the material phase and on the electromagnetic phase. The information contained in the Maxwell equations is examined in Appendix 2.

**Table 2 : Main balances of practical use on material and electromagnetic phases with source terms transferred between them.**

| | *Material Phase* | *Electromagnetic Phase* |
|---|---|---|
| ***Mass*** | $\dfrac{\partial \rho_{(e)}}{\partial t} = -\operatorname{div} \boldsymbol{J}$ | – |
| ***Momentum*** | $\dfrac{\partial}{\partial t}\left[\rho \boldsymbol{v}\right] = -\dfrac{\partial}{\partial x_i}\left[\rho v_i v_j + \Pi_{ij}\right] + \boldsymbol{F}$ <br> alternative expressions of the ponderomotive force $\boldsymbol{F}$ are given in [14,15,26]. | [Minkowski definition) <br> $\dfrac{\partial(\boldsymbol{D}\times\boldsymbol{B})}{\partial t} = \dfrac{\partial}{\partial x_i} T_{ij} - \boldsymbol{F}$ or <br> $\dfrac{\partial(\boldsymbol{D}\times\boldsymbol{B})}{\partial t} = \dfrac{\partial}{\partial x_i}\left[D_i E_j + B_i H_j - (\tfrac{1}{2}\varepsilon_0 E^2 + \tfrac{1}{2}\mu_0 H^2)\delta_{ij}\right]$ <br> $-\mathcal{P}_j \dfrac{\partial E_j}{\partial x_i} - \mathcal{J}_j \dfrac{\partial H_j}{\partial x_i} - \rho_e \boldsymbol{E} - \boldsymbol{J}\times\boldsymbol{B}$ <br> where $T_{ij}$ is the Maxwell stress tensor |
| ***Total Energy*** <br><br> *Electrodynamics of Continua Applications* | $\dfrac{\partial}{\partial t}\left[\dfrac{1}{2}\rho v^2 + \rho \dot{U}\right] = -\operatorname{div}[\dfrac{1}{2}\rho v^2 \boldsymbol{v} + \tau_{ij}\times\boldsymbol{v} + \rho \dot{U}\boldsymbol{v}$ <br> $+ P\boldsymbol{v} - (\boldsymbol{E}\cdot\boldsymbol{\mathcal{P}})\boldsymbol{v} - (\boldsymbol{H}\cdot\boldsymbol{\mathcal{J}})\boldsymbol{v} + \boldsymbol{J}_Q] + \boldsymbol{E}\cdot\boldsymbol{J}$ | Poynting equation : <br> $\dfrac{\partial w}{\partial t} = -\operatorname{div}\boldsymbol{S} - \boldsymbol{E}\cdot\boldsymbol{J}$ |
| *Radiative Transfer Applications* | $\dfrac{\partial}{\partial t}\left[\dfrac{1}{2}\rho v^2 + \rho\dot{U} + \rho_{(e)}\dot{\Phi}\right] = -\operatorname{div}[\dfrac{1}{2}\rho v^2\boldsymbol{v} + \rho_{(e)}\dot{\Phi}\boldsymbol{v}$ <br> $+ \Pi_{ij}\times\boldsymbol{v} + \rho\dot{U}\boldsymbol{v} + \sum_k \dot{\Phi}_k \boldsymbol{J}_k + \boldsymbol{J}_Q] + \mathcal{A} - \mathcal{E}$ | $\dfrac{\partial w}{\partial t} = -\operatorname{div}\boldsymbol{J}_R - \mathcal{A} + \mathcal{E}$ |

**Constitutive and Wave Equations for Linear and Isotropic Media**

As already discussed, the two independent pairs of homogeneous and non-homogeneous Maxwell equations revealed different status. It clearly appears from balance eq. 17 that the vector source $J^\lambda$ (charges and electric currents) produces, independently of the given medium, a 4-tensor of electric and magnetic excitations $G^{\lambda\mu}$, whereas electromagnetic forces are linked to the 4-tensor for fields $F_{\lambda\mu}$. In order to have a relation between excitations and fields, depending of the electromagnetic properties of the material phase (the so-called closure problem), it is necessary to define a constitutive equation. For a linear medium, a constitutive tensor $\chi$ is introduced giving (see Appendix 1):

$$G^{\lambda\nu} = \frac{1}{2}\chi^{\lambda\nu\mu\rho}F_{\mu\rho} \quad (26)$$

Formally, from the symmetry relations applying to the constitutive tensor $\chi$, only 20 independent components are necessary to characterize a given medium [10], but for linear and isotropic media, corresponding to many practical chemical engineering operations, the general four dimensional tensorial equation (26) reduces to the classical pair of constitutive equations:

$$\boldsymbol{D} = \varepsilon_r\varepsilon_0\boldsymbol{E} \quad (27)$$

$$\boldsymbol{H}\,\mu_r\mu_0 = \boldsymbol{B} \quad (28)$$

The permittivities and permeabilities involved in eq. (27-28) are defined in Table 3. These constitutive equations enable then to define the polarization $\mathcal{P}$ and magnetization $\mathcal{M}$ vectors expressing the electric and magnetic properties of a considered medium from:

$$\mathcal{P} = \boldsymbol{D} - \varepsilon_0\boldsymbol{E} \quad (29)$$

$$\mathcal{M} = \frac{\boldsymbol{B}}{\mu_0} - \boldsymbol{H} \quad (30)$$

These vector fields have been previously shown to play a crucial role in the definitions of the local electric and magnetic moments, which are the main extensive quantities involved in the generalized definitions of enthalpy and Gibbs equation for electromagnetic systems [14].

In the same manner, it is possible to use the constitutive equations (27-28) to define any electromagnetic state from only two vectors fields. Generally, in order to perform an energetic treatment of the problem involving the Poynting vector, the ***E*** and ***H*** vectors are retained. It is then easy to combine Maxwell equations (18-19, 22-23) with the constitutive equations (27-28) to obtain the so-called wave equations for a linear, isotropic medium verifying the electroneutrality [13]:

$$\Delta\boldsymbol{E} + \gamma^2\boldsymbol{E} = 0 \quad (31)$$

$$\Delta\boldsymbol{H} + \gamma^2\boldsymbol{H} = 0 \quad (32)$$

where $\gamma^2 = \omega^2\varepsilon\mu$ .

These wave equations are a starting point to solve any problem of electromagnetic wave propagation in conducting or dielectric media, as a preliminary step to the formulation of the coupling with any chemical engineering application. In all cases, the electromagnetic behavior of the system is strongly related to the electromagnetic properties of the considered material phase. The reciprocal relations between electromagnetic properties and optical properties of any linear and isotropic non-magnetized medium are given in Table 3.

**<u>Table 3 :</u> Reciprocal relations between electromagnetic and optical properties for a linear, isotropic and conducting medium (for a simple dielectric medium, the imaginary part of complex quantities vanishes).**

| ***Electromagnetic Properties*** | ***Optical Properties*** |
|---|---|
| Permittivity of free space $\varepsilon_0 = 8.85418781762\ 10^{-12}$ F.m$^{-1}$<br>Permeability of free space $\mu_0 = 4\pi\ 10^{-7}$ H.m$^{-1}$<br>Relative permeability $\mu_r = \frac{\mu}{\mu_0}$ | |
| Complex relative permittivity (dielectric constants)<br>$\varepsilon_r = \varepsilon_r^{'} - i\,\varepsilon_r^{''}$ | Complex refractive index<br>$m = n - i\ \kappa = \sqrt{\varepsilon_r \mu_r}$ |
| Ohm's law: $\boldsymbol{J} = \sigma \boldsymbol{E}$<br>$\sigma = \frac{\lambda_0 c_0 \mu}{4\pi\, n\, \kappa}$ Conductivity | |
| ***Reciprocal relations for a non-magnetized medium with $\mu = \mu_0$*** | |
| $\varepsilon_r^{'} = n^2 - \kappa^2 = \varepsilon \mu c_0^2 = \frac{\varepsilon^{'}}{\varepsilon_0}$<br>$\varepsilon_r^{''} = 2n\kappa = \frac{\varepsilon^{''}}{\varepsilon_0} = \frac{\sigma}{\omega \varepsilon_0}$ | $n = \sqrt{\frac{\sqrt{\varepsilon_r^{'2} + \varepsilon_r^{''2}} + \varepsilon_r^{'}}{2}}$<br>$\kappa = \sqrt{\frac{\sqrt{\varepsilon_r^{'2} + \varepsilon_r^{''2}} - \varepsilon_r^{'}}{2}}$ |

**Classification of Chemical Engineering Problems Related to Electrodynamics and Strategy for Modeling**

As explained in introduction, the number of applications for electrodynamics in chemical engineering processes is considerable in numerous domains, both in conventional and high technology industry. Examining the formal structure of the balance equations and formulating their coupling may authorize to propose a classification of these problems, even if it is always a partial and difficult task. Two main classes of problems can be addressed, defining a strong or a weak coupling between electromagnetic field and the material phase.

***Electrodynamics of Continua***

In this case, strong interactions appear in the material phase put in the electromagnetic field. The medium may be charged or not, and conduction or convection currents may exist. All the main balances, including mass and momentum balances, need to be reformulate [20]. New source terms appear, corresponding to long range and non-conservative forces acting on the medium and requiring to be clearly defined (Tables 1 and 2). Of course, these chemical engineering applications are non-relativistic and the quasi-static form of Maxwell's equations are excellent approximations. The considered quasi-static field for each case then serves to define different classes in chemical engineering applications (Table 4).

***Wave Transport Phenomena***

These applications involve all the problems of electromagnetic waves propagation in complex media. Because the coupling with the material phase only exists through the energy balance, it is called a weak coupling. Two different ways are then possible to treat the relevant problems, depending of the case study. First, the electromagnetic wave approach consists to solve the Maxwell equations in the form of a wave equation (eq. 31-32). This leads to the spatial distribution of the electric and magnetic fields in the medium. This approach is preferred for non-scattering media (conducting media), and when the penetration length of the wave is lower or of the same order of magnitude than the characteristic length of the process, as it is often the case for micro-wave processes. Second, the radiative transfer approach [21, 22] enables to calculate the local irradiance field in scattering and complex media. The two methods lead to local variables calculations, related to important integral quantities, as for example the mean volumetric rate of electromagnetic energy transferred between the material and the photonic phase, which can induce thermal or chemical modifications of the material phase [23]. It must be emphasized that, even in the radiative formulation, any practical use of balances requires to know the optical properties of the material medium, i.e. the absorption and scattering coefficients and the phase function for scattering. These coefficients can be theoretically calculated from the elemental electromagnetic properties of the medium (Table 3) using the Lorenz-Mie theory, which corresponds to the computation of the Maxwell equations on "particles" of different shapes and sizes [24, 25].

***Other Classification Criteria***

The two previous classes of applications still involve many different problems and additional criteria exist, examining the mathematical nature of the considered application. It is first important to define if the problem is linear or not, in regard to the electrodynamics, i.e. if the material medium may interact with the electromagnetic fields. This appears each time that convection currents exist or when the medium is charged, which considerably increases the mathematical treatment complexity. A simplification of great importance is also to discuss if the material medium may be considered as non-polarizable (electrically or magnetically). Finally, it is necessary to distinguish if the problem is "direct" (actual modeling of a phenomenon) or "inverse" (model inversion from indirect observations of the phenomenon).

The applications in electrodynamics of continua are generally direct problems, but many are non-linear. In all cases, the mathematical treatment is simplified for the non-polarizable medium hypothesis. For the applications in wave transport phenomena, only the linear problem presents some interest in chemical engineering. It corresponds to the study of free waves propagation in complex media. In these cases, one can distinguish between direct modeling problems and inverse problems, these latter being often used in sophisticated analytical techniques relying on light and matter interactions.

The different classes discussed in this section with a non-exhaustive list of examples and applications in the field of chemical engineering [20] are summarized in Table 4.

**Table 4 : Electrodynamics and Chemical Engineering Applications (proposal for problems classification)**

| **ELECTRODYNAMICS OF CONTINUA**<br>**(polarizable or non-polarizable medium)** | | **WAVE TRANSPORT PHENOMENA**<br>**(linear case = free wave)** | |
|---|---|---|---|
| ***LINEAR***<br>(convection current free and $\rho_e \equiv 0$ ) | ***NON LINEAR***<br>(convection currents and $\rho_e \neq 0$ or = 0) | ***DIRECT*** | ***INVERSE*** |
| - Electrochemical processes (Electrostatics)<br><br>- Joule effect Heating<br><br>- Induction Heating<br><br>- Electric and Magnetic fields separations<br><br>(Electrophoresis, Electro-phoretic Centrifugation)<br><br>- Galvanomagnetic Effects<br><br>- Thermomagnetic Effects | Magnetohydrodynamics (MHD = quasi-static magnetic field):<br><br>- Charged media (plasmas, ionized media...)<br>- Liquid metals<br>- Nuclear reactors<br><br>Electrohydrodynamics (EHD = quasi-static electric field):<br><br>- Pumping and levitation of liquids and gases<br>- Gas electrodynamic compression<br>- poly-electrolytes (proteins, ADN...)<br>- Enhancement of heat transfer<br>- Extraction of contaminants from gases<br>- Property measurements in fluid systems<br><br>Ferrofluids:<br><br>- Magnetic stabilization of fluidized beds<br>- Lubricants and withstand of highly corrosive gases<br>- Magnetocaloric energy conversion<br>- Fluidmagnetic buoyancy<br>- Magnetic-fluid pumping<br>- Flow separations | Radiative Transfer:<br><br>- High temperature processes<br>- Combustion<br>- Radiation heating<br>- Infra-red processes<br>- Radiative equilibrium<br>- Solar energy<br><br>____________<br><br>- Chemical Photo-reactors and<br>- Photobioreactors<br><br>- UV applications<br><br>- High frequency heating<br><br>- Micro-wave applications (reactors, drying, heating...)<br><br>- Optical and Radiative properties determination (absorption, scattering, phase function,...)<br><br>- Ionizing radiations (X, γ)<br><br>- Laser applications | - Particle characterization (micro-organisms, mean diameter, shape, bubble sizing.......)<br><br>- Granular and porous media characterization<br><br>- Hyperfrequency sensor<br><br>- Transmitance and reflectance applications (turbidimetry,.....)<br><br>- Particles RTD<br><br>- Retro-diffusion Laser |

**Concluding Remarks**

In order to generalize the classical balances formulation to chemical engineering applications dealing with electrodynamics, it is necessary to use a covariant formalism in a four dimensions space-time continuum. This enables to distinguish between balances and field equations, and to focus on the main extensive quantities involves in electromagnetic phenomena. This latter point is of crucial importance in combining balances of conserved and non-conserved quantities for a generalized thermo-mechanics treatment of an arbitrary continuum [15].

Examining the formal structure of such covariant equations also authorizes to discuss the theoretical foundations and the underlying determinism in modeling electrodynamics applications for chemical engineering. By the same way, in defining the strength and the linearity of the coupling between the material and the electromagnetic phases, one can select different classes of problems relevant to electrodynamics of continua, or to wave transport phenomena.

The practical use of intensive field equations always requires to know the constitutive relations for the material phase (permittivity, permeability,…or derived “radiative” quantities as absorption and scattering coefficients, phase function for scattering,…), related to the molecular properties of the considered matter [16]. Moreover, even if it was not considered in this paper, it generally requires to develop the total energy balance from a thermodynamic treatment using for example the Gibbs equation [3, 14-15].

Although the current presentation is limited to local aspects because electrodynamics problems are intrinsically local, a special attention must be further paid on volumetric integration of local balances [20], leading to engineering equations of practical use [4, 16].

Finally, a more rigorous general treatment, unifying dynamics and electrodynamics balances in relation with chemical engineering applications still remains to formulate in the near future. In this paper, we have restricted the analysis to the classical eulerian approach, which is not the most suitable in order to establish the theoretical foundations of chemical engineering problems with electromagnetic fields. This latter objective will be probably reached using a covariant hamiltonian approach [13] in connection to the generalized bracket formulation [27-28], enabling to include in the formalism, other important classes of balances for chemical engineering applications such as populations balances or Boltzmann-type balances.

# NOTATIONS

| | | |
|---|---|---|
| $\mathcal{A}$ | Volumetric rate of radiant energy absorption | [W.m$^{-3}$] |
| $\boldsymbol{A}$ | Magnetic potential vector | [Wb.m$^{-1}$] |
| $\boldsymbol{B}$ | Magnetic induction | [T] |
| $c$ | Velocity of light | [m.s$^{-1}$] |
| $\boldsymbol{D}$ | Electric induction | [C.m$^{-2}$] |
| $\mathcal{E}$ | Volumetric rate of radiant energy emission | [W.m$^{-3}$] |
| $E_{TOT}$ | Volumetric density of total energy | [J.m$^{-3}$] |
| $\boldsymbol{E}$ | Electric field | [V.m$^{-1}$] |
| $\boldsymbol{F}$ | Ponderomotive force by unit volume | [N.m$^{-3}$] |
| $F_{\lambda\mu}$ | 4-pseudotensor for fields (six-vector fields) | |
| $\boldsymbol{g}$ | Gravitational acceleration | [m.s$^{-2}$] |
| $G^{\lambda\mu}$ | 4-pseudotensor for excitations (six-vector excitations) | |
| $\boldsymbol{H}$ | Magnetic field | [A.m$^{-1}$] |
| $\boldsymbol{j}$ | Conductive electric current density | [A.m$^{-2}$] |
| $\boldsymbol{J}$ | Total current density | [A.m$^{-2}$] |
| $\boldsymbol{J}_{E_{TOT}}$ | Total energy flux density | [W.m$^{-2}$] |
| $J_{(ij)}$ | Flux density of $\Psi$ | [$\Psi$ .m$^{-2}$.s$^{-1}$] |
| $\boldsymbol{J}_Q$ | Thermal conductive flux density | [W.m$^{-2}$] |
| $\boldsymbol{J}_R$ | Radiant energy flux density | [W.m$^{-2}$] |
| $\boldsymbol{J}_v$ | Convective electric current density | [A.m$^{-2}$] |
| $J^{\mu}$ | 4-vector of flux density | |
| $J^{\lambda\mu}$ | 4-tensor of flux density | |
| $\mathcal{J}$ | Magnetic polarization | [T] |
| $\mathcal{L}$ | Lagrangian density | [J.m$^{-3}$] |
| $m$ | Complex refractive index | [-] |
| $\boldsymbol{m}$ | Local magnetic moment | [A.m$^{2}$] |
| $\mathcal{M}$ | Magnetization | [A.m$^{-1}$] |
| $n$ | Real part of the refractive index | [-] |
| $\boldsymbol{p}$ | Local electric moment | [C.m] |
| P | Pressure | [Pa] |
| $\mathcal{P}$ | Electric polarization | [C.m$^{-2}$] |
| Q | Charge | [C] |
| S | Surface | [m$^{2}$] |
| $\boldsymbol{S}$ | Poynting vector | [W.m$^{-2}$] |
| t | Time | [s] |
| $T_{ij}$ | Maxwell stress tensor | [Pa] |
| $\dot{U}$ | Mass internal energy density | [J.kg$^{-1}$] |
| $\boldsymbol{v}$ | Velocity | [m.s$^{-1}$] |
| V | Volume | [m$^{3}$] |
| V | Electric potential | [V] |
| w | Maxwell electromagnetic energy density | [J.m$^{-3}$] |
| $w_m$ | Mass flow rate | [kg.s$^{-1}$] |
| $x_i$ | Cartesian coordinates | [m] |
| $x^{\lambda}$ | Lorentz coordinates | |

***Greek Letters***

| | |
|---|---|
| $\delta_{ij}$ | Kronecker delta [0] |
| $\varepsilon$ | Permittivity [F.m$^{-1}$] |
| $\varepsilon_r$ | Relative permittivity [-] |
| $\eta$ | Dynamic viscosity [Pa.s] |
| $\kappa$ | Imaginary part of the refractive index (absorption) [-] |
| $\lambda$ | Wavelength [m] |
| $\mu$ | Permeability [H.m$^{-1}$] |
| $\mu_r$ | Relative permeability [-] |
| $\Pi_{ij}$ | Total pressure tensor [Pa] |
| $\rho$ | Density [kg.m$^{-3}$] |
| $\rho_e$ | Total charge density [C.m$^{-3}$] |
| $\sigma_{(i)}$ | Source of $\Psi_{(i)}$ [$\Psi$.m$^{-3}$.s$^{-1}$] |
| $\sigma$ | Conductivity [S.m$^{-1}$] |
| $\tau_{ij}$ | Shear stress [Pa] |
| $\dot{\Phi}$ | Potential energy mass density [J.kg$^{-1}$ ou V.m] |
| $\dot{\Phi}_k$ | Potential energy mass density for the component k [J.kg$^{-1}$ ou V.m] |
| $\chi^{\lambda\mu\nu\rho}$ | 4-tensor for medium electromagnetic properties |
| $\psi_{(i)}$ | Volumetric density of $\Psi_{(i)}$ [$\Psi$.m$^{-3}$] |
| $\Psi_{(i)}$ | Extensive quantity [USI] |
| $\dot{\Psi}_{(i)}$ | Mass density of $\Psi_{(i)}$ [$\Psi$.kg$^{-1}$] |
| $\omega$ | Pulsation [s$^{-1}$] |

***Superscripts***

| | |
|---|---|
| ' | Relative to the real part of the complex permittivity |
| '' | Relative to the imaginary part of the complex permittivitty |

***Subscripts***

| | |
|---|---|
| c | Relative to conduction |
| f | Relative to the moving frontier F |
| l | Relative to convection |
| 0 | Relative to free space properties |

# APPENDIX 1

## Proof of the Electromagnetic Moment Covariant Balance (eq. 17) from the Least Action Principle

The least action principle [2 , 9, 11] is a variational method widely used in classical mechanics to establish momentum equations. It can be used also in tensorial field theory, and generalized to continua by defining a Lagragian density $\pounds$. It then appears as a powerful tool and a concurrent method in demonstrating balances, specially in case of electrodynamics when the field potentials variation is required [2, 11].

Considering the general relation between the 4-tensor of fields $F$ and the 4-tensor of excitations $G$ from the tensor of electromagnetic properties for the medium $\chi$ [10] $G^{\lambda\nu} = \frac{1}{2}\chi^{\lambda\nu\mu\rho}F_{\mu\rho}$, it is possible to write the Lagrangian density for the electromagnetic field as:

$$\pounds = \frac{1}{8}\chi^{\lambda\nu\mu\rho}F_{\lambda\nu}F_{\mu\rho} - J^{\lambda}A_{\lambda} \qquad \text{(A1)}$$

where $\boldsymbol{A}$ is the 4-vector generalized potential.

The action for the system is then defined by $S = \int_{t_1}^{t_2} \pounds \, d\Omega$ where $d\Omega = dt\,dV$ , leading from eq. (A1) to :

$$S = -\int\left(\frac{1}{8}\chi^{\lambda\nu\mu\rho}F_{\lambda\nu}F_{\mu\rho} - J^{\lambda}A_{\lambda}\right)d\Omega = -\int\left(\frac{1}{4}G^{\lambda\nu}F_{\lambda\nu} - J^{\lambda}A_{\lambda}\right)d\Omega \qquad \text{(A2)}$$

Applying the least action principle for the variation of the integral, and taking into account that the current must not vary [11], it comes:

$$\delta S = -\int\left[-J^{\lambda}\delta A_{\lambda} + \frac{1}{8}\left(\chi^{\lambda\nu\mu\rho}F_{\lambda\nu}\delta F_{\mu\rho} + \chi^{\lambda\nu\mu\rho}F_{\mu\rho}\delta F_{\lambda\nu}\right)\right]d\Omega = 0 \qquad \text{(A3)}$$

the two last terms being equal, one obtains:

$$\delta S = -\int\left[-J^{\lambda}\delta A_{\lambda} + \frac{1}{4}\left(\chi^{\lambda\nu\mu\rho}F_{\mu\rho}\delta F_{\lambda\nu}\right)\right]d\Omega = -\int\left[-J^{\lambda}\delta A_{\lambda} + \frac{1}{2}G^{\lambda\nu}\delta F_{\lambda\nu}\right]d\Omega = 0 \qquad \text{(A4)}$$

and from the definition of $F_{\lambda\nu} = \frac{\partial A_{\nu}}{\partial x^{\lambda}} - \frac{\partial A_{\lambda}}{\partial x^{\nu}}$ :

$$\delta S = -\int\left[-J^{\lambda}\delta A_{\lambda} + \frac{1}{2}G^{\lambda\nu}\left(\frac{\partial}{\partial x^{\lambda}}\delta A_{\nu} - \frac{\partial}{\partial x^{\nu}}\delta A_{\lambda}\right)\right]d\Omega = 0 \qquad \text{(A5)}$$

which leads, by permutation of subscripts $\lambda$ and $\nu$ on which the summation is performed and taking $F_{\nu\lambda} = - F_{\lambda\nu}$:

$$\delta S = -\int\left[-J^{\lambda}\delta A_{\lambda} + \frac{1}{2}G^{\lambda\nu}\frac{\partial}{\partial x^{\nu}}\delta A_{\lambda}\right]d\Omega = 0 \qquad \text{(A6)}$$

Integrating by part the second integral of eq. (A6), or applying the Gauss theorem one has :

$$\delta S = -\int\left[-J^{\lambda} + \frac{\partial G^{\lambda\nu}}{\partial x^{\nu}}\right]\delta A_{\lambda}d\Omega - \int G^{\lambda\nu}\delta A_{\lambda}dS_{\nu}\Big| \qquad \text{(A7)}$$

In the second term, the integral must be evaluated to the limits of integration. Considering the coordinates, these limits are infinity because the field vanishes at infinity. About the initial and final given times, one can consider that there is no potentials variation, because, from the least action principle, the potentials are known at these times. Consequently, the second term of eq. (A7) vanishes, leading to:

$$\delta S = -\int\left[-J^{\lambda} + \frac{\partial G^{\lambda\nu}}{\partial x^{\nu}}\right]\delta A_{\lambda}d\Omega = 0 \qquad \text{(A8)}$$

because the variations $\delta A_{\lambda}$ were arbitrary, the coefficient for $\delta A_{\lambda}$ must be equal to zero, i.e.:

$$\frac{\partial G^{\lambda\nu}}{\partial x^{\nu}} = J^{\lambda} \qquad \text{(A9)}$$

which correspond to the balance equation for extensive quantity of electromagnetic field (eq. 17).
An analogous proof, based on the hamiltonian formalism is available in the textbook of Panofsky and Phillips [12].

## APPENDIX 2

### Self-content of Maxwell equations

———————————————

The Maxwell equations correspond to a set of two non-homogeneous equations for electromagnetic moments balance:

$$\text{div}\,\boldsymbol{D} = \rho_e \quad \text{(A1)}$$

$$\text{curl}\boldsymbol{H} - \frac{\partial \boldsymbol{D}}{\partial t} = \boldsymbol{J} \quad \text{(A2)}$$

and a set of two homogeneous equations describing the non-stationary fields:

$$\text{div}\,\boldsymbol{B} = 0 \quad \text{(A3)}$$

$$\text{curl}\,\boldsymbol{E} + \frac{\partial \boldsymbol{B}}{\partial t} = 0 \quad \text{(A4)}$$

Taking the divergence of equation (A2) and applying the identity $\text{div curl} \equiv 0$ , one obtains:

$$\text{div curl}\,\boldsymbol{H} - div\frac{\partial \boldsymbol{D}}{\partial t} = div\boldsymbol{J} \text{ , i.e. } \text{div}\frac{\partial \boldsymbol{D}}{\partial t} + div\boldsymbol{J} = 0$$

and with the help of eq. (A1):

$$\text{div}\left(\frac{\partial \boldsymbol{D}}{\partial t}\right) = \frac{\partial}{\partial t}(\text{div}\boldsymbol{D}) = \frac{\partial \rho_e}{\partial t} \text{ which corresponds to } -\text{div}\,\boldsymbol{J} = \frac{\partial \rho_e}{\partial t} \quad \text{(A5)}$$

This equation corresponds to the charge conservation (eq. 15) and then appears to be already contained in the first pair of Maxwell equations.